\documentclass[11pt]{article}

\usepackage[pdftex]{graphicx}

\usepackage{hyperref}
\pdfoutput=1
\begin{document}
\title{Evaporation in motion}
\author{Hatim Machrafi, Alexey Rednikov, Pierre Colinet, Pierre C. Dauby \\
\\\vspace{6pt} Institute of Physics, Universite de Liege, Belgium \\
TIPs, Fluid Physics Unit, Universite Libre de Bruxelles, Belgium} 
\maketitle
\begin{abstract}
This work presents fluid dynamics videos obtained via numerical (CFD) calculations using ComSol (finite elements method) software,
showing the evaporation of HFE7100 (3M company refrigerant) into a nitrogen gas flow along the liquid interface. The overall temperature evolution and liquid motion, which is caused by surface-tension (Marangoni) and buoyancy (Rayleigh) instability mechanisms, are shown as well. Flow behavior in the liquid caused by the aforementioned instability mechanisms can be nicely seen. Finally, these observations are made for three liquid thicknesses in order to appreciate the qualitative influence of confinement. 
\end{abstract}
\section{Introduction}
\begin{figure}
\resizebox{0.8\columnwidth}{!}{\includegraphics{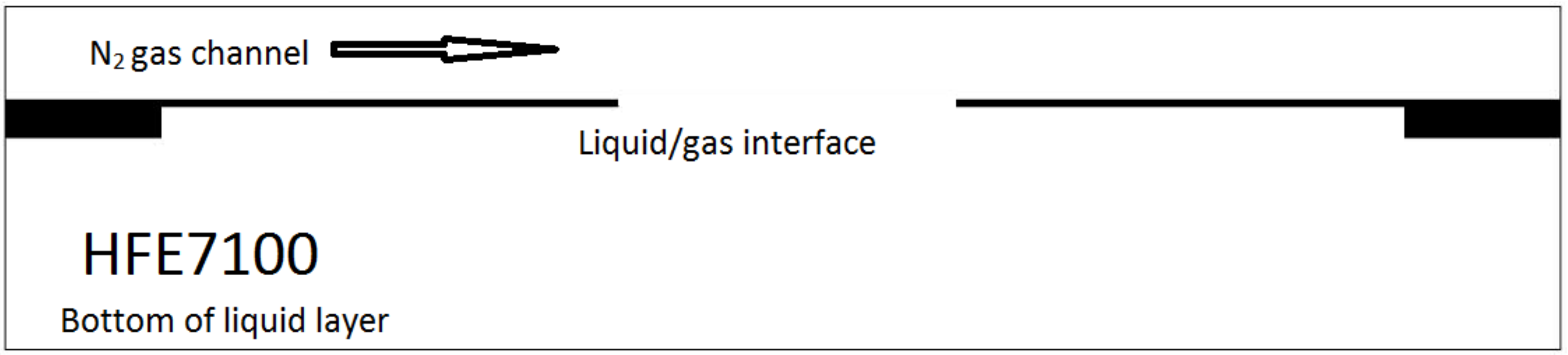} } 
\caption{A scheme of the system}
\label{Scheme}
\end{figure}

This work falls into a general framework which consists of observing the behavior of patterns and structures that can be formed after instability onset in an evaporating liquid layer. In previous work, we studied theoretical instability thresholds in pure fluids
[1,2] and in binary mixtures [3,4]. What is of interest here, is a two-dimensional numerical simulation study of the transient temperature and fluid motion in the liquid for a liquid evaporating into a nitrogen gas flow. The chosen liquid is HFE7100 (an electronic liquid produced by 3M). The numerical (CFD) simulations are performed using the software ComSol (finite elements method). The evaporation causes the instability and the gas flow evacuates the liquid vapor. The setup used for this numerical simulation is represented in Fig. \ref{Scheme} and is inspired from the CIMEX experimental setup of ESA [5]. 

The gas flow is maintained at 100 ml/min in a channel of 3 mm height, while three different liquid thicknesses are considered: 2, 4 and 8 mm. The width of the whole setup is 50 mm. The cover between the liquid and gas channel is $200 \mu m$ thick. At the middle of this cover, there is an opening with a width of 10.6 mm, allowing contact between the liquid and gas channel. These items are taken into the geometry of the numerical software ComSol. The boundaries of the whole system are kept at an ambient temperature and pressure of respectively 298 K and 1 atm, except for the gas channel outlet where only the ambient pressure is respected. Also, the whole system is surrounded by walls except for the gas flow inlet and outlet. The interface is kept at a constant height, since in the ESA experimental setup the liquid is to be replenished at the same rate as the evaporation rate. At the interface, flux conservation is maintained and a tangential stress balance is considered. Furthermore, a no-slip condition is assumed at the interface. The assumption of local thermodynamic equilibrium at the interface allows us the use of Raoult's law, in which the temperature dependence of the saturation pressure is determined via the Clausius-Clapeyron relation. 

The results present the temperature in the liquid and gas phase as well as the fluid motion in the liquid (caused by the evolution of the temperature via surface-tension and buoyancy effects) by means of streamlines as a function of time. The real total elapsed time is 10 seconds.
Two videos are shown, presenting the same results in the following URLs: 
\begin{enumerate}
\item \href{DOI}{Video 1 - High resolution} 
\item \href{DOI}{Video 2 - Low resolution}
\end{enumerate}
Note that in the videos the inner streamlines represent the highest velocity values. The red color represents the highest observed temperature (that of the ambient one), 298 K. The blue color represents the lowest observed temperature, around 285 K.

\section{Discussion}
\begin{figure}
\resizebox{0.3\columnwidth}{!}{\includegraphics{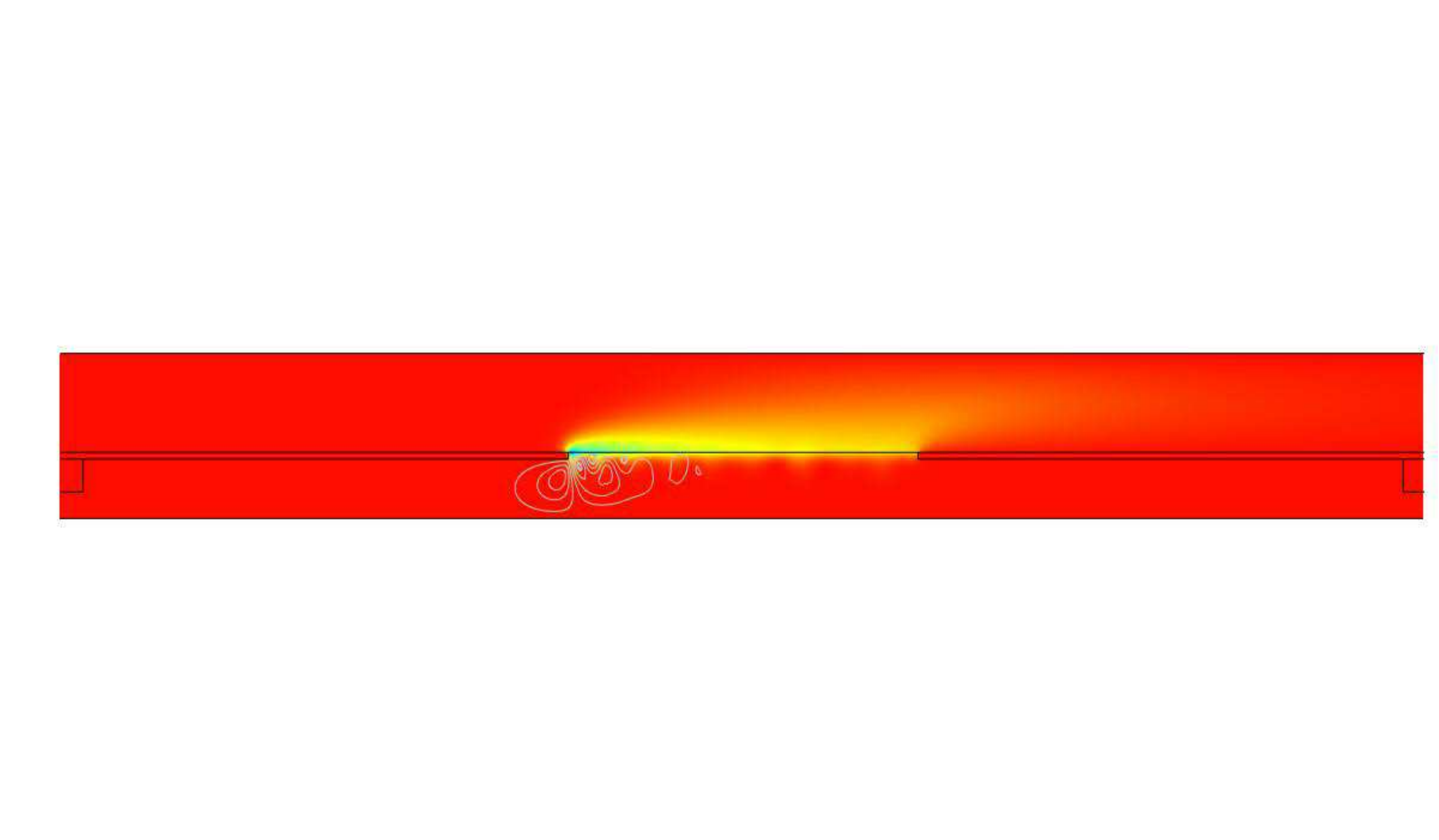} } 
\quad\resizebox{0.3\columnwidth}{!}{\includegraphics{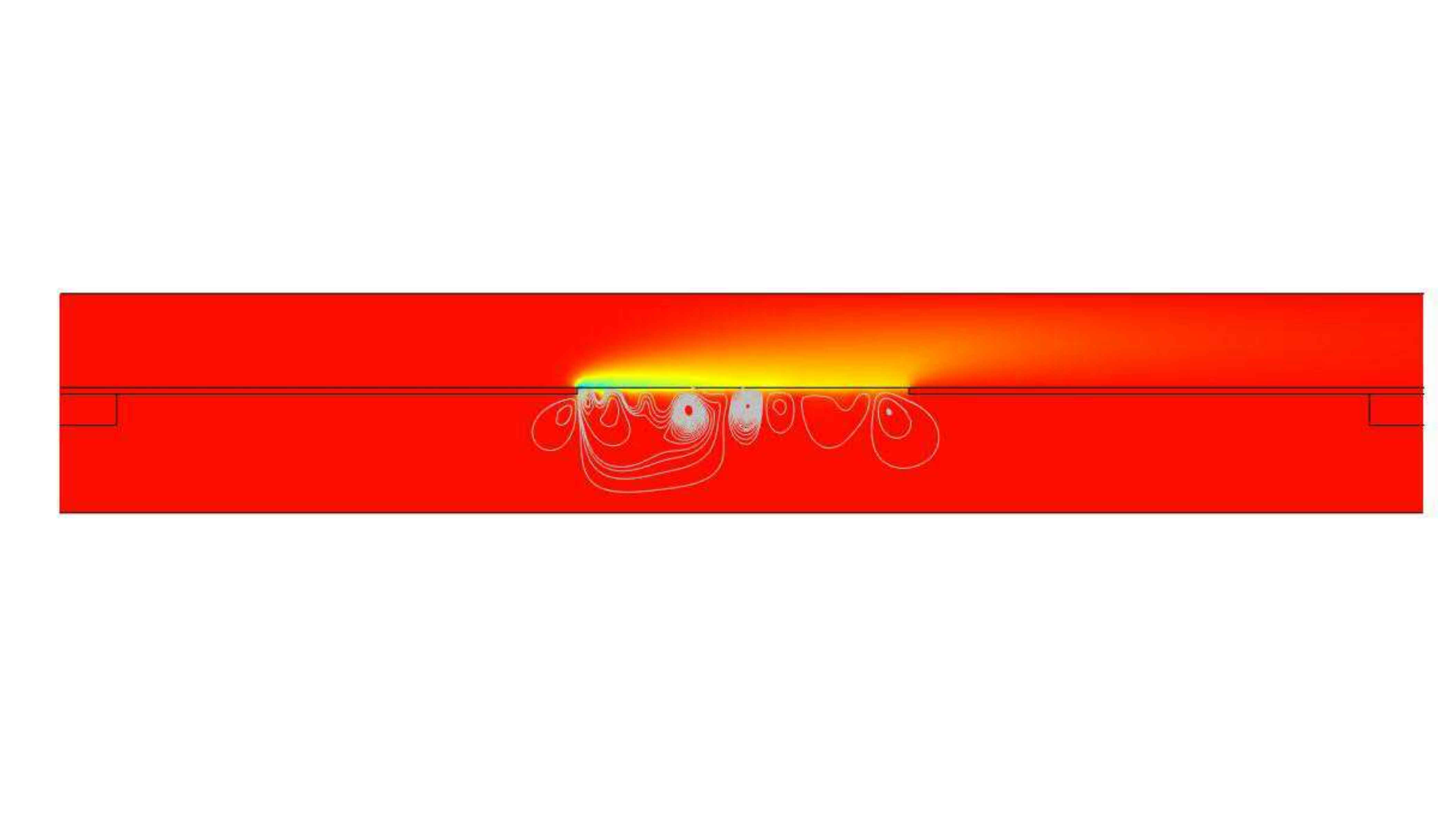} }
\quad\resizebox{0.3\columnwidth}{!}{\includegraphics{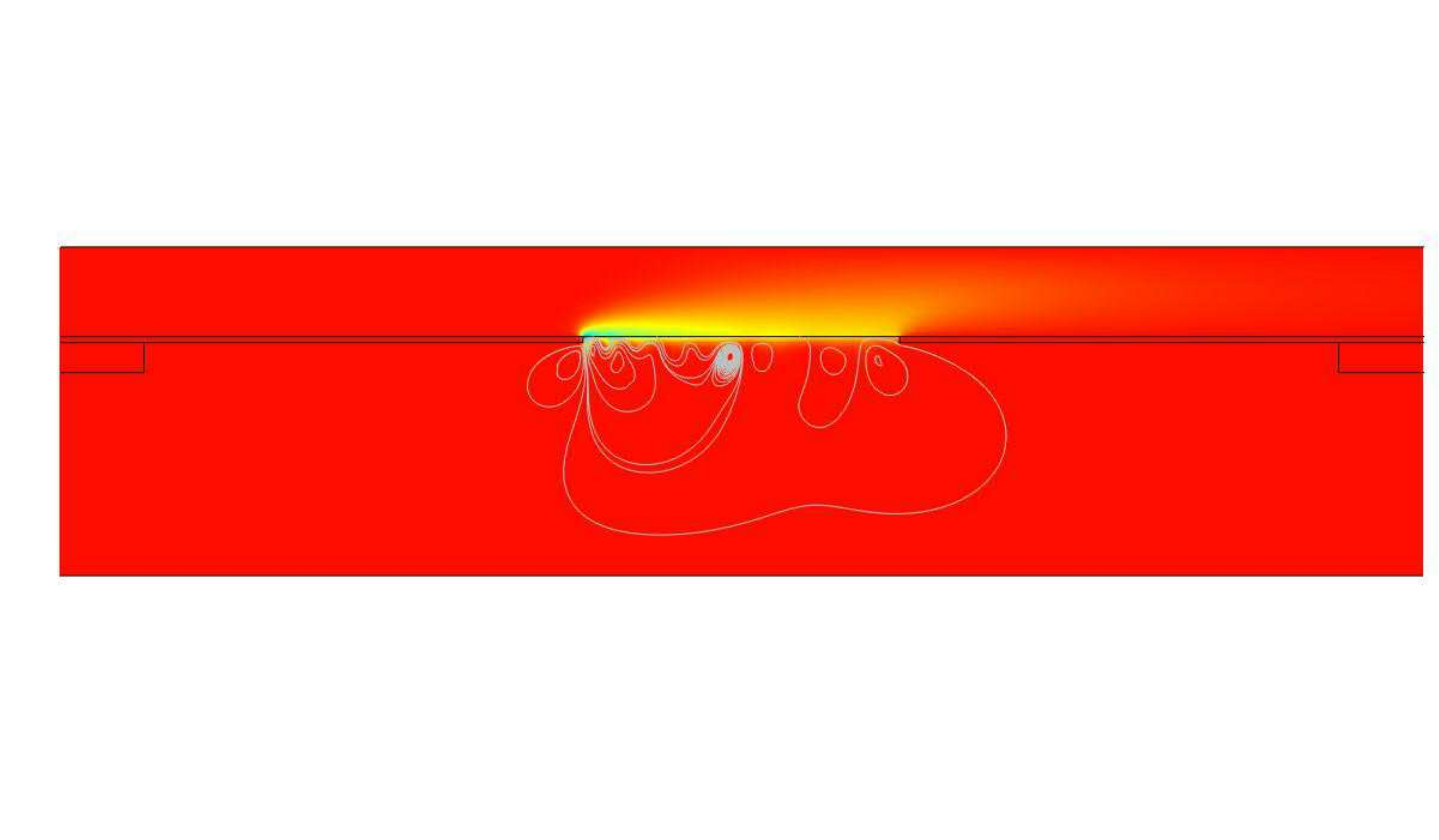} }
\caption{The temperature and liquid flow pattern at time $t=1 s$ for liquid thicknesses of 2 mm (left), 4 mm (middle) and 8 mm (right) and a gas flow of 100 ml/min}
\label{comparisont1}
\end{figure}

From the results in the videos we can observe that first several small rolls are formed near the surface, caused by the surface-tension effect as Fig. \ref{comparisont1} shows for the three liquid layer thicknesses at time $t=1 s$.

\begin{figure}
\resizebox{0.3\columnwidth}{!}{\includegraphics{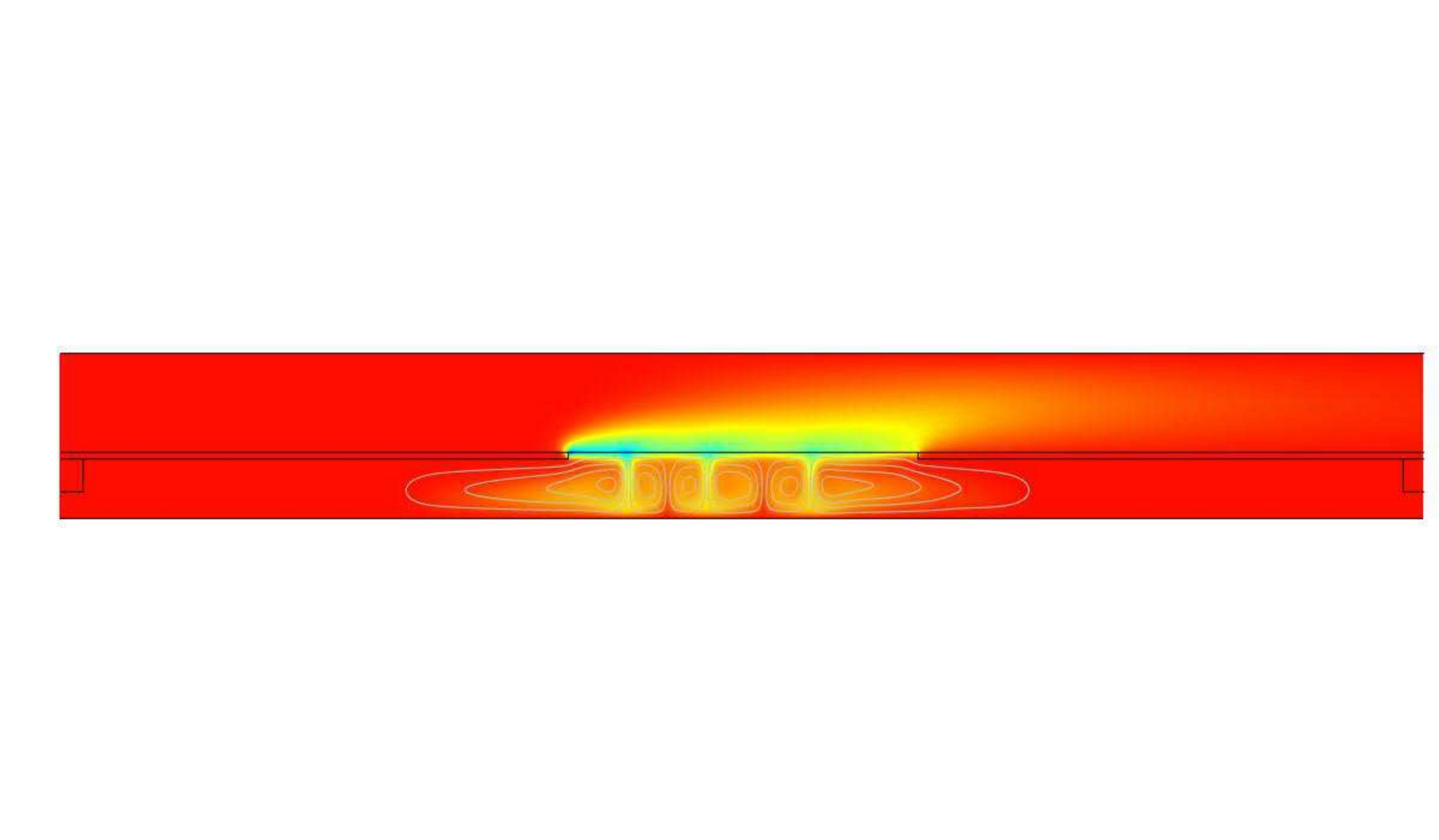} } 
\quad\resizebox{0.3\columnwidth}{!}{\includegraphics{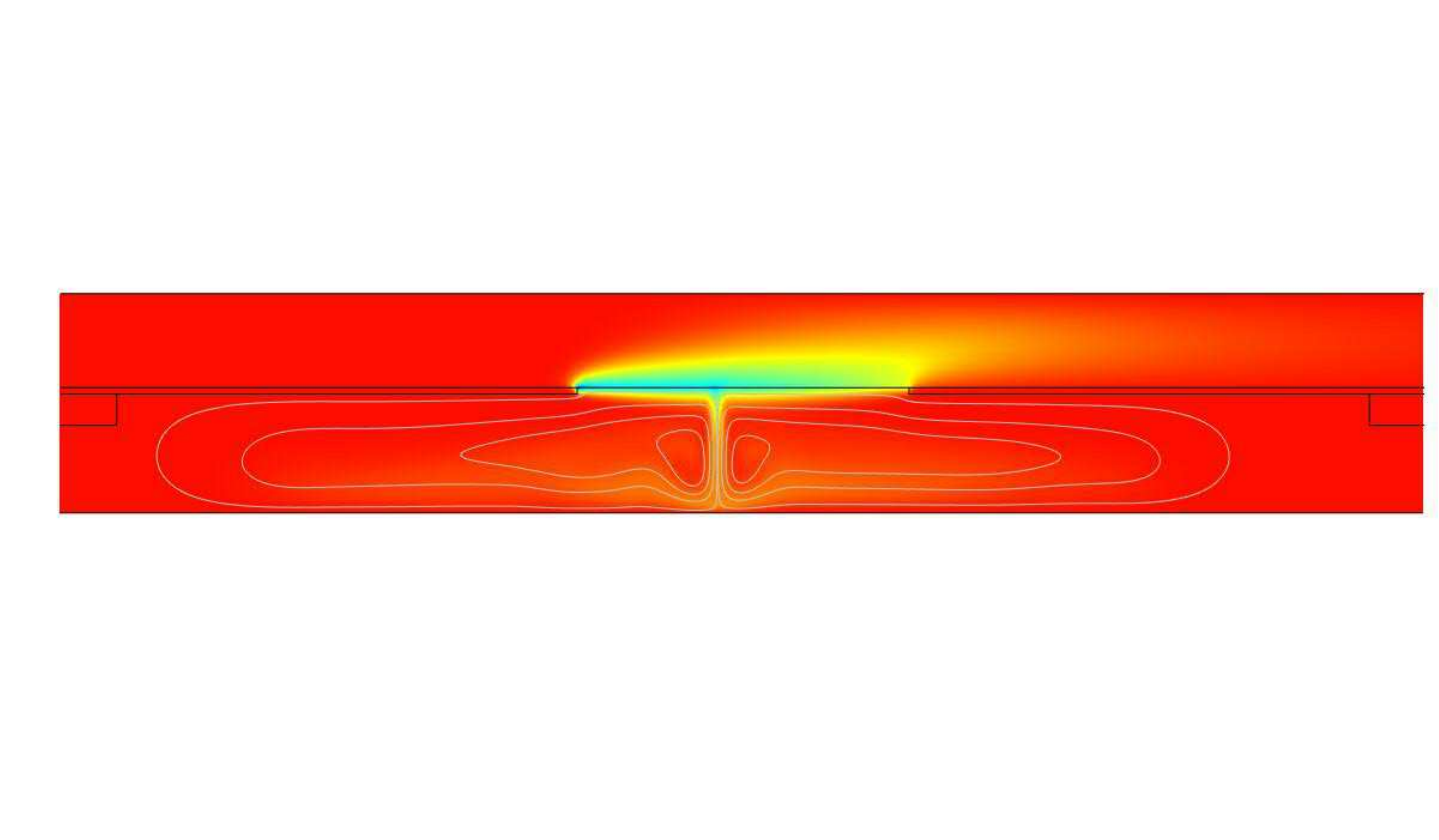} }
\quad\resizebox{0.3\columnwidth}{!}{\includegraphics{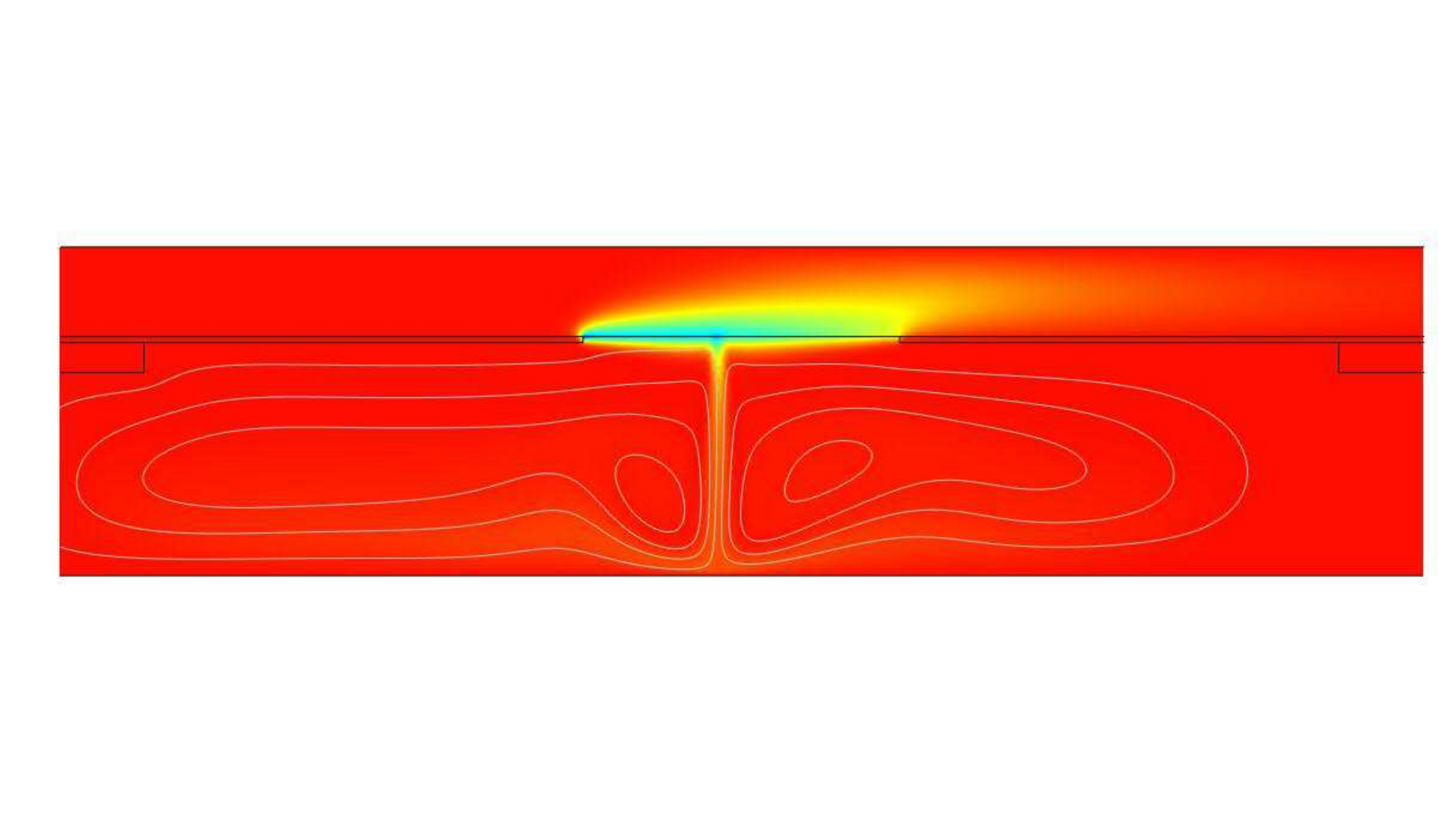} }
\caption{The temperature and liquid flow pattern at time $t=10 s$ for liquid thicknesses of 2 mm (left), 4 mm (middle) and 8 mm (right) and a gas flow of 100 ml/min}
\label{comparisont10}
\end{figure}

\noindent Due to buoyancy and as time proceeds, the rolls grow towards the bottom of the liquid layer. Then the rolls also grow in horizontal direction merging with each other until a steady configuration is obtained. For a higher liquid layer thickness, the merging occurs earlier and less rolls are left. Furthermore, the temperature gradients decrease as the liquid thickness increases, which is caused by the higher mixing efficiency when the liquid is less confined. Moreover, the rolls extend more horizontally under the cover towards the side walls as the liquid layer thickness increases. For smaller liquid layer thicknesses, the rolls reach the bottom where a constant temperature of 298 K is maintained. Therefore the rolls stay concentrated close to the interface. As the liquid layer thickness increases, the rolls have more time to increase in size towards the side walls before they reach the bottom of the liquid layer. Fig. \ref{comparisont10} shows this at the time $t= 10 s$.

\noindent This work yields valuable information about the supercritical instability behavior of an evaporating liquid and the qualitative influence of its confinement by means of fluid dynamics.

\section{Acknowledgments}
The authors gratefully acknowledge financial support of BelSPo and ESA.

\section{References}
[1] B. Haut and P. Colinet, J. Colloid Interface Sci., 285: 296-305, 2005.
[2] F. Chauvet, S. Dehaeck and P. Colinet, Europhys. Lett., 99: 34001, 2012.
[3] H. Machrafi, A. Rednikov, P. Colinet, P.C. Dauby, J. Colloid Interface Sci., 349: 331-353, 2010.
[4] H. Machrafi, A. Rednikov, P. Colinet, P.C. Dauby, Eur. Phys. J., 192: 71-81, 2011.
[5] ESA, \href{http://www.esa.int/SPECIALS/HSF_Research/SEMLVK0YDUF_0.html}{CIMEX experimental setup}, accessed 12 octobre 2012

\end{document}